\documentclass[ reprint,superscriptaddress,
,nofootinbib, amsmath,amssymb, floatfix, aps, tight ]{revtex4-2}
\usepackage{algorithm}
\usepackage[noend]{algpseudocode} % pseudocode style for algorithmicx
\usepackage{graphicx}% Include figure files
\usepackage{dcolumn}% Align table columns on decimal point
\usepackage{bm}% bold math
\usepackage[colorlinks=true,linkcolor=teal,citecolor=teal]{hyperref}

\usepackage[mathlines]{lineno}% Enable numbering of text and display math

\usepackage{csquotes}
\usepackage{multirow} 
\usepackage{amsmath}
\usepackage[bb=pazo]{mathalpha}
\usepackage{mathrsfs} 
\usepackage{braket}
\usepackage{xcolor}
\usepackage{bbm}
\usepackage{comment}

\usepackage{enumitem} % for customizing lists
\usepackage{siunitx}
\usepackage{times}

\definecolor{darkgreen}{rgb}{0,0.5,0}

\definecolor{newblue}{rgb}{0.0, 0.5, 0.69}

\begin{document}

\title{High-rate multipartite quantum secret sharing with composable security}

\author{Russell M.~J.~Brooks}
\altaffiliation{These authors contributed equally.}
\affiliation{Institute of Photonics and Quantum Sciences, School of Engineering and Physical Sciences, Heriot-Watt University, Edinburgh, United Kingdom}
\author{Joseph Ho}
\altaffiliation{These authors contributed equally.}
\affiliation{Institute of Photonics and Quantum Sciences, School of Engineering and Physical Sciences, Heriot-Watt University, Edinburgh, United Kingdom}
\author{Joseph Niblo}
\affiliation{Institute of Photonics and Quantum Sciences, School of Engineering and Physical Sciences, Heriot-Watt University, Edinburgh, United Kingdom}
\author{Janka Memmen}
\affiliation{Electrical Engineering and Computer Science Department, Technische Universität Berlin, Berlin, Germany}
\author{Anna Pappa}
\affiliation{Electrical Engineering and Computer Science Department, Technische Universität Berlin, Berlin, Germany}
\author{Jens Eisert}
\affiliation{Dahlem Center for Complex Quantum Systems, Freie Universität Berlin, Berlin, Germany}
\author{Nathan Walk}
\affiliation{Dahlem Center for Complex Quantum Systems, Freie Universität Berlin, Berlin, Germany}
\author{Alessandro Fedrizzi}
\email{A.Fedrizzi@hw.ac.uk}
\affiliation{Institute of Photonics and Quantum Sciences, School of Engineering and Physical Sciences, Heriot-Watt University, Edinburgh, United Kingdom}

\date{\today}

\begin{abstract}
Future quantum communication networks will conceivably support cryptographic tasks that require entanglement among more than two users. Quantum secret sharing is a prime example where entanglement provides a direct means to coordinate untrusted parties with security from eavesdropping in a multi-party setting. However, the canonical GHZ-based protocols can be vulnerable to participant attacks, in which untrusted parties try to learn the secret without collaborating. 
Here, we experimentally evaluate a discrete-variable $(n,n)$-threshold quantum secret-sharing protocol whose finite-key analysis provides composable security against general attacks, including participant attacks.
Using two domain-engineered entangled photon pair sources, we generate 4-qubit GHZ states at rates above $5\times10^3$ fourfold events per second and a maximum asymptotic secret key rate of $750 \pm 10$ bits per second. We then distribute the state through a 4-arm star network comprising 20~km of fibre in total. 
From the measured event rates and error statistics, we infer that a randomised 24-hour execution with the optimised basis probability would yield a composable finite-key lower bound of 8.7 Mbits, under the assumption that the measured source and device statistics remain stationary.

\end{abstract}

\maketitle

\section{\label{intro}Introduction}

Secret sharing is a cryptographic primitive that distributes information among $n$ parties such that only authorised subsets of at least $k$ parties, with $k\leq n$, can reconstruct the original secret, while any unauthorised subset learns nothing about it. Originally introduced as a mechanism for distributed storage and protection~\cite{shamir1979share,blakley1979safeguarding}, classical schemes such as Shamir’s threshold secret-sharing scheme~\cite{shamir1979share} encode a secret into multiple shares that can be held at different locations or by different custodians. This enhances robustness against data loss while preserving confidentiality against the compromise of fewer than $k$ shares.
The compelling idea of \emph{quantum secret sharing} (QSS) extends this concept using quantum-information-processing techniques, which can additionally reveal attempts at eavesdropping or unauthorised access~\cite{hillery1999quantum}. Beyond the sharing of classical information, QSS can distribute an unknown quantum state such that only authorised groups can reconstruct it, a task known as quantum state sharing~\cite{cleve1999share,Gottesman:2000vx,lu2016secret}. These capabilities make QSS a natural primitive for secure multiparty computation~\cite{cirac1999distributed,crepeau2002secure,ouyang2017computing}, controlled access to information~\cite{zhang2006controlled,singh2024controlled}, and distributed quantum-information storage~\cite{Gottesman:2000vx,cleve1999share}.

The first explicit proposal of QSS was introduced by Hillery, Bu\v{z}ek and Berthiaume~\cite{hillery1999quantum} (HBB), and independently conceived by Zukowski et al,~\cite{Zukowski:1998}, who showed that multipartite entanglement correlations can be used to securely distribute shares of a classical secret.
In the HBB protocol, one of the participants, called the dealer, prepares an $(n+1)$-partite GHZ~\cite{greenberger1990bell} state and distributes one subsystem to each participant.
By performing multiparty measurements in randomly chosen bases and publicly announcing a subset of results, the dealer verifies the shared quantum resource before using it to encode the secret.
This combines distribution and security testing in a single step, rather than requiring separate \emph{quantum key distribution} (QKD)~\cite{bennett1984proceedings} links between the dealer and each participant~\cite{fujiwara2016unbreakable,richter2021agile,liu2023experimental}. 
Crucially, this marks a fundamental step beyond point-to-point quantum communication: rather than treating a quantum network merely as a collection of pairwise links, QSS exploits genuinely multipartite correlations as a native resource for networked cryptographic tasks.
Furthermore, the multipartite correlations are advantageous in a fully connected quantum network~\cite{kimble2008quantum, wehner2018quantum} where they provide more efficient use of network resources~\cite{Epping_2017,pant2019routing}.
Related multipartite cryptographic primitives include conference key agreement and its anonymous variants, in which a subset of network users establishes a common key while concealing the participants' identities~\cite{hahn2020anonymous,Hahn2023}.

The original HBB construction realises an $(n,n)$-threshold scheme, in which all $n$ shares are required to reconstruct the secret, reflecting the fully symmetric correlations of the GHZ state~\cite{greenberger1990bell}.
Subsequent work extended this framework to general $(k,n)$-threshold schemes, analogous to Shamir’s construction, by employing graph states whose correlation structures provide more flexible access types~\cite{hein2004multiparty,markham2008graph}. QSS has since been demonstrated using discrete variable GHZ and graph states~\cite{tittel2001experimental,chen2005experimental,gaertner2007experimental,hwang2011multiparty, bell2014experimental}, continuous variable multipartite states~\cite{lance2004tripartite,armstrong2015multipartite,Cai:2017cp}, and single qubit schemes~\cite{schmid2005experimental,Grice2019}.

However, these demonstrations have been subject to so-called participant attack strategies~\cite{karlsson1999quantum, xiao2004efficient, qin2007cryptanalysis, Hoi_Lo2007}.
In the original HBB protocol, all participants are involved in the certification process which involves public announcements to certify the correlations.
A dishonest participant can exploit going last to ensure they can always pass these rounds.
Two partial solutions have been proposed.
Williams \textit{et al}~\cite{williams2019quantum} addressed this by randomising the order in which participants respond to the dealer, forcing a dishonest member to make a measurement announcement first, at least some of the time.
Kogias \textit{et al}~\cite{kogias2017unconditional} instead treated every unauthorised subset as part of the adversary model and proved security against the worst coalition.
Both of these strategies are valid in the asymptotic limit of infinite rounds, and the security proofs effectively reduce to QKD analysis~\cite{walk2021sharing}.
However, Ref.~\cite{williams2019quantum} uses a uniform measurement basis, which is not optimal, even in the asymptotic regime, and Ref.~\cite{kogias2017unconditional} considered only continuous variable states without finite-key effects and, applied to the HBB protocol, strictly yields negative rates.

Recently, composably secure $(n,n)$-QSS protocols have been developed that resolve this challenge in both the continuous variable~\cite{walk2021sharing} and discrete variable regimes~\cite{memmen2023advantageQSS}.
The key insight was to make use of a measurement scheme that allowed pairwise correlations to be established between the dealer and each participant.
This allowed one to bound the information leaked using entropic uncertainty relations of Ref.~\cite{kogias2017unconditional}.
Here, we implement and evaluate the discrete-variable $(n,n)$-QSS protocol~\cite{memmen2023advantageQSS} using photonic GHZ states.
By taking advantage of high-purity entangled photon pair sources with four-fold generation exceeding $5\times 10^3$ per second, we infer an asymptotic extractable secret-key rate of $750\pm 10$ bits/s.

\section{\label{protocol}Secret sharing protocol}

\begin{figure}[t]
    \centering
    \includegraphics[width=\linewidth]{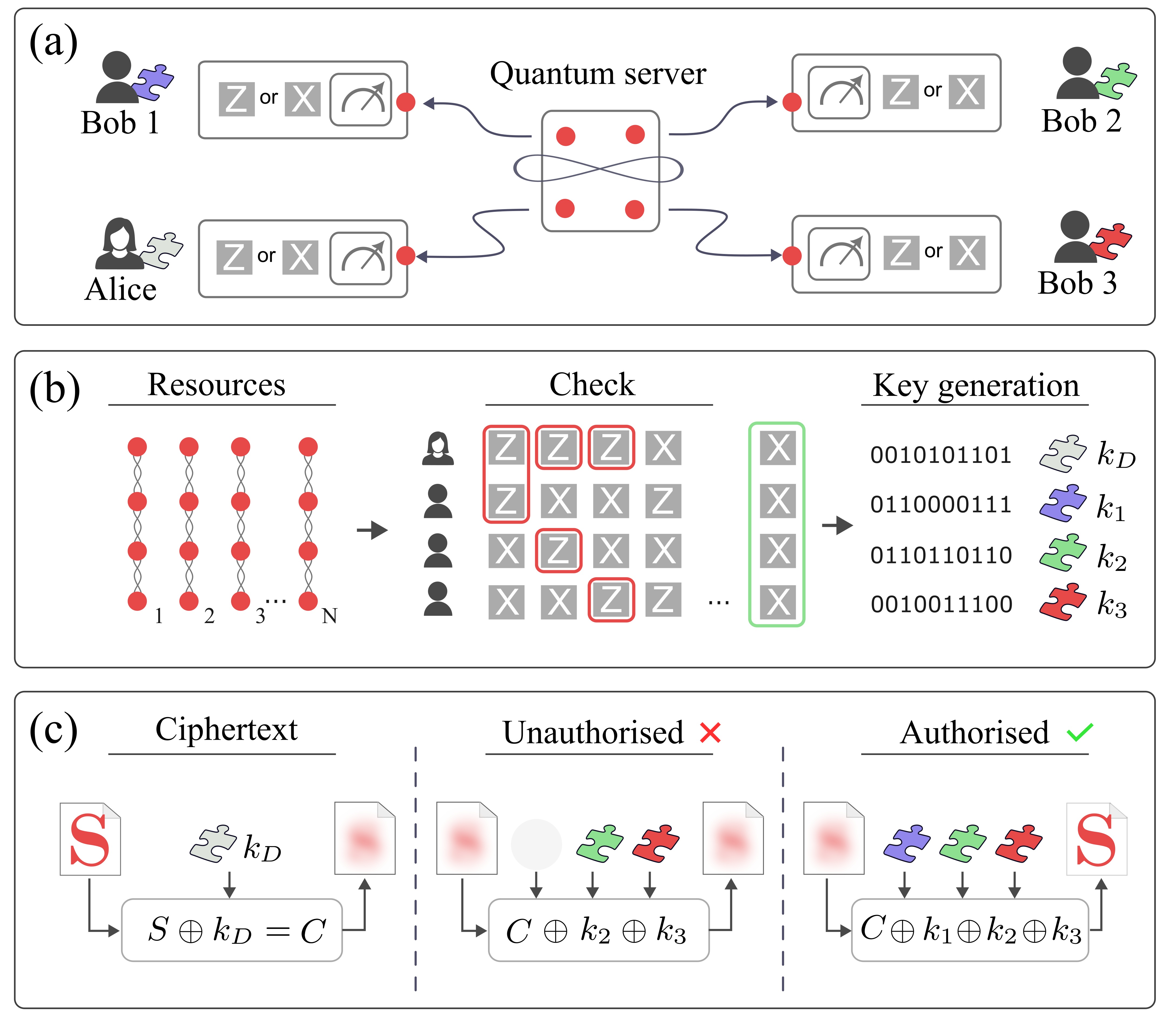}
    \caption{\textbf{Quantum secret sharing protocol}.
    (a) Alice (dealer) requests a server to deliver $N$ copies of the GHZ state to all users (Bobs) over quantum channels. Each user performs measurements in $Z$  or $X$  basis. 
    (b) Once all $N$ GHZ states are measured, the dealer 
    performs basis reconciliation and identifies rounds in which they and at least one other user measure in the $Z$  basis in red boxes. These check rounds are used for parameter estimation, while joint measurements in $X$  are shown in the green box and are for key generation. After error correction and privacy amplification, each party obtains one share of the total secure key, labelled $k_i$.
    (c) The dealer uses their share, $k_D$ to encrypt a secret message, $S$ and publicly broadcasts the cipher.
    To decrypt the cypher and reveal the secret, \textit{all} participants must combine their individual keys to form an authorised set.}
    \label{fig:concept}
\end{figure}

In this work we consider an $(n,n)$-quantum secret sharing scheme, where the secret is encoded into $n$ shares, all of which are required to recover the secret.
We briefly outline the protocol, which is shown conceptually in Fig.~\ref{fig:concept}.
In each round, an $n+1$ qubit GHZ state is distributed between the dealer Alice and the participants Bob-$i$, $i\in\{1, \dots, n\}$.
Each user performs a measurement on their qubit, either in the Pauli-$Z$ or -$X$ basis, chosen at random.
After many rounds, they reveal their basis choices and basis reconciliation is performed.
Rounds where all users measured in $X$ are used for \textit{key-generation}, since in the ideal case, the measurement outcomes are perfectly correlated and can be used to form a raw secret key.
The quantum bit error rate in these rounds is $Q_X = (1-\braket{X^{\otimes n+1}})/2$.
Rounds where Alice and at least one Bob measured in $Z$ are used for \textit{check} rounds.
From these, Alice computes pairwise correlations with each party as $Q_{A,B_i}=(1-\braket{Z_AZ_{B_i}})/2$, which bound the information accessible to unauthorised subsets of users or an external eavesdropper.
The third type of rounds contains all unused permutations which are discarded.
Following Ref.~\cite{memmen2023advantageQSS}, the users can bias their basis choices to maximise the fraction of key-generation rounds and therefore the secret key rate, without compromising security from a participant attack.
In the asymptotic limit with infinite rounds, the fractional key rate after privacy amplification and error correction is given by
\begin{equation}\label{eq_akr}
    \mathrm{AKR} = 1-h_2(Q_X)-\underset{B_i\ \in \mathcal{B}}{\mathrm{max}}h_2(Q_{A,B_i} ),
\end{equation}
where $h_2(x)$ is the binary Shannon entropy.
For complete details on the security definitions for correctness and secrecy, see Ref.~\cite{memmen2023advantageQSS}.

\section{\label{results}Experiment}
 
\begin{figure}
    \centering
    \includegraphics[width=\linewidth]{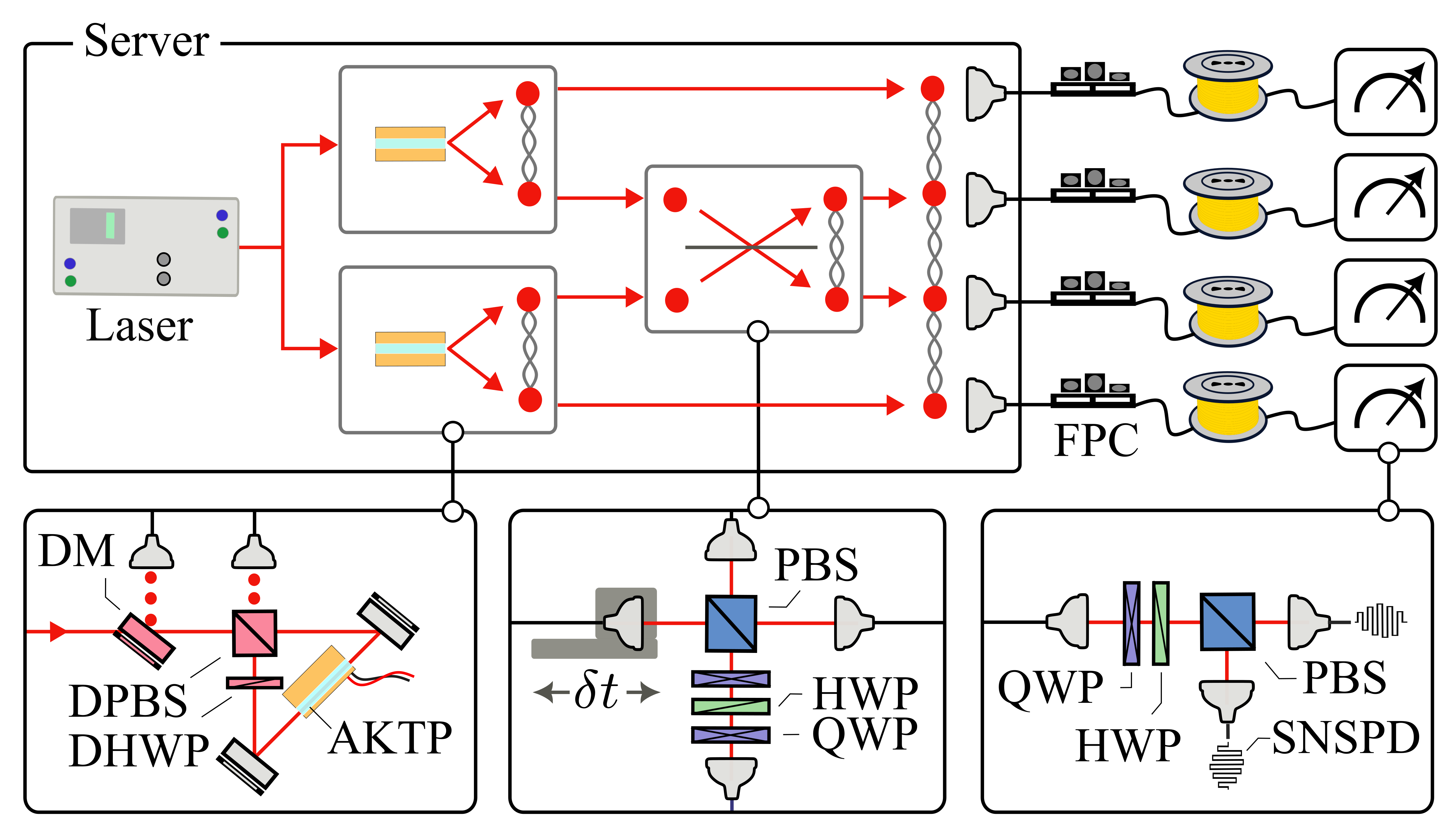}
    \caption{\textbf{Experiment setup}. The quantum server uses a pulsed laser that pumps the two entangled photon pair sources. Each source uses an aperiodically-poled KTP crystal embedded in a Sagnac interferometer made from a \emph{dichroic mirror} (DM), \emph{dichroic polarising beam splitter} 
    (DPBS), and a dichroic half-wave plate (DHWP). One photon from each source is sent to a PBS which probabilistically generates the four-photon GHZ state. Each photonic qubit is sent through a \emph{fibre polarisation controller} 
    (FPC) and fibre spool to a party, who are equipped with polariser analysers and superconducting nanowire \emph{single-photon detectors} (SNSPDs).}
    \label{fig:exp_layout}
\end{figure}

We experimentally prepare a four-photon GHZ state using a linear optics setup as shown in Fig.~\ref{fig:exp_layout}.
Two \emph{entangled photon-pair sources} 
(EPS) are pumped by a 775~nm picosecond pulsed laser to produce photon pairs at 1550~nm, via \emph{spontaneous parametric down-conversion} (SPDC).
Each EPS use aperiodically-poled potassium titanyl phosphate (AKTP) crystal, which is engineered to create spectrally pure biphoton states for enhancing multi-photon interference~\cite{Graffitti:18, pickston2021optimised}.
The AKTP crystal is embedded in a Sagnac interferometer to generate polarisation entanglement~\cite{fedrizzi2007wavelength}.
One photon from each EPS interferes on a polarising beamsplitter, where they become entangled in the polarisation degree of freedom. The entangling gate succeeds with a probability of one-half~\cite{Rudolph_fusion}, and conditioned on one photon exiting each PBS port, we post-select onto a four-qubit GHZ state vector
\begin{equation}
    |\textrm{GHZ}\rangle= \frac{|h,h,h,h\rangle+e^{i \phi} |v,v,v,v\rangle}{\sqrt{2}},
\end{equation}
where $|h\rangle$ is the horizontal and $|v\rangle$ the vertical polarisation state vector and $\phi$ is the relative phase nominally set to zero.
Each photon in the setup is coupled into a single-mode fibre and sent to polarisation measurement stages. 
The GHZ state is verified using quantum state tomography to reconstruct the density matrix. 
We observe a maximum generation rate of 5025 (counts/s) with fidelity $0.842 \pm 0.002$, see Appendix~\ref{supp_tomo} for details.

The QSS protocol was implemented first in the asymptotic regime with zero transmission loss to benchmark the source performance. We measure the GHZ state using sixteen settings corresponding to the full set of joint $Z$  and $X$  basis measurements performed by the four users.
The AKR is evaluated using the expression in Eq.~(\ref{eq_akr}).
Using the measured four-fold rate $r$, we evaluate the overall secure key rate as $\mathrm{SKR} := AKR \times r$.
We measured the SKR over a range of pump powers, starting from 100~mW up to 680~mW to explore the impact of errors created by multi-photon terms and the expected performance of the resultant key rate.
We measured the maximum SKR to be $750\pm10$ (bits/s) at 600~mW pump power. In this regime, the source operates with an average four-fold rate of $r=3970$~(counts/s) and the corresponding noise parameters, $Q_X=0.082\pm0.001$ and $\textrm{max}[{Q_{A,B_i}}]=0.0808\pm0.0004$, yielding an average AKR of $0.188\pm0.003$, see Appendix~\ref{app_optimal} for further details.

Using the optimal source power, we experimentally evaluate the QSS protocol in a fibre-network emulation comprising 20 km of fibre in total.
The source is placed in the middle of a network in a star topology with 5~km links connecting each user; see Appendix~\ref{fibre_net} for more information. The total loss was approximately 5~dB, resulting in a four-fold generation rate $r=1197$ per second. Each user measures symmetrically in the 
 $Z$  and $X$  bases for 10 hours of laboratory time and collects 22 million rounds for processing. Approximately half the measurement time is used switching basis. The average error rates are $Q_X=0.090 \pm 0.003$ and $\mathrm{max}[ Q_{A,B_i}]=0.075 \pm 0.001$. 

We apply the finite-key analysis of Ref.~\cite{memmen2023advantageQSS} to network emulation data to account for the finite-size sampling effects. The secure finite-size key length $\ell$ is given by
\begin{equation} \label{fin_key}
\begin{aligned}
\ell \ge {} &  m(1-\varepsilon_{rob}) \Bigg[ 1 - f_{EC} h_2 \Bigl( Q_X + \xi_1  \Bigr) 
\\
&
-  h_2 \Bigl( \max_i Q^{k_i}_{A  B_i} + \xi_2 \Bigr) \Bigg]
-\log_2 \Biggl( \frac{n}
{2\varepsilon_{\mathrm{EC}}\varepsilon_{\mathrm{PA}}^2} \Biggr),
\end{aligned}
\end{equation}
where $m$ are the total number of key generation rounds, and $k_i$ denotes the check rounds in the subset $i$ and $n=3$ is the number of participants excluding the dealer. 
The various $\varepsilon$ terms are related to the total epsilon security, which is set at $\varepsilon_{\mathrm{total}}=10^{-10}$, see table~\ref{table_params} in Appendix~\ref{supp_finite}.
The function $\xi_1(\varepsilon,m):=\sqrt{\mathrm{log}(1/\varepsilon)/m}$ is the Hoeffding bound~\cite{hoeffding1963probability} and $\xi_2(\varepsilon,m,k):=\sqrt{(m+k)(k+1)\mathrm{ln}(1/\varepsilon)/mk^2}$ is the Serfling bound~\cite{serfling1974probability}, used to capture the finite sample statistics. We estimate an error correction inefficiency of $f_{EC}=1.1$ using efficient Polar codes~\cite{liao2025efficient}. The security definition of Eq.~(\ref{fin_key}) allows us to bias the key generation measurement basis with probability $p_{\rm key}$, and optimise the key length; see Appendix~\ref{supp_finite} for details. We model the secure key length in post-processing using both the symmetric basis measurement and the optimised biased basis probability $p_{\rm key}^*$. 

\begin{figure}[t]
    \centering
    \includegraphics[width=1\linewidth]{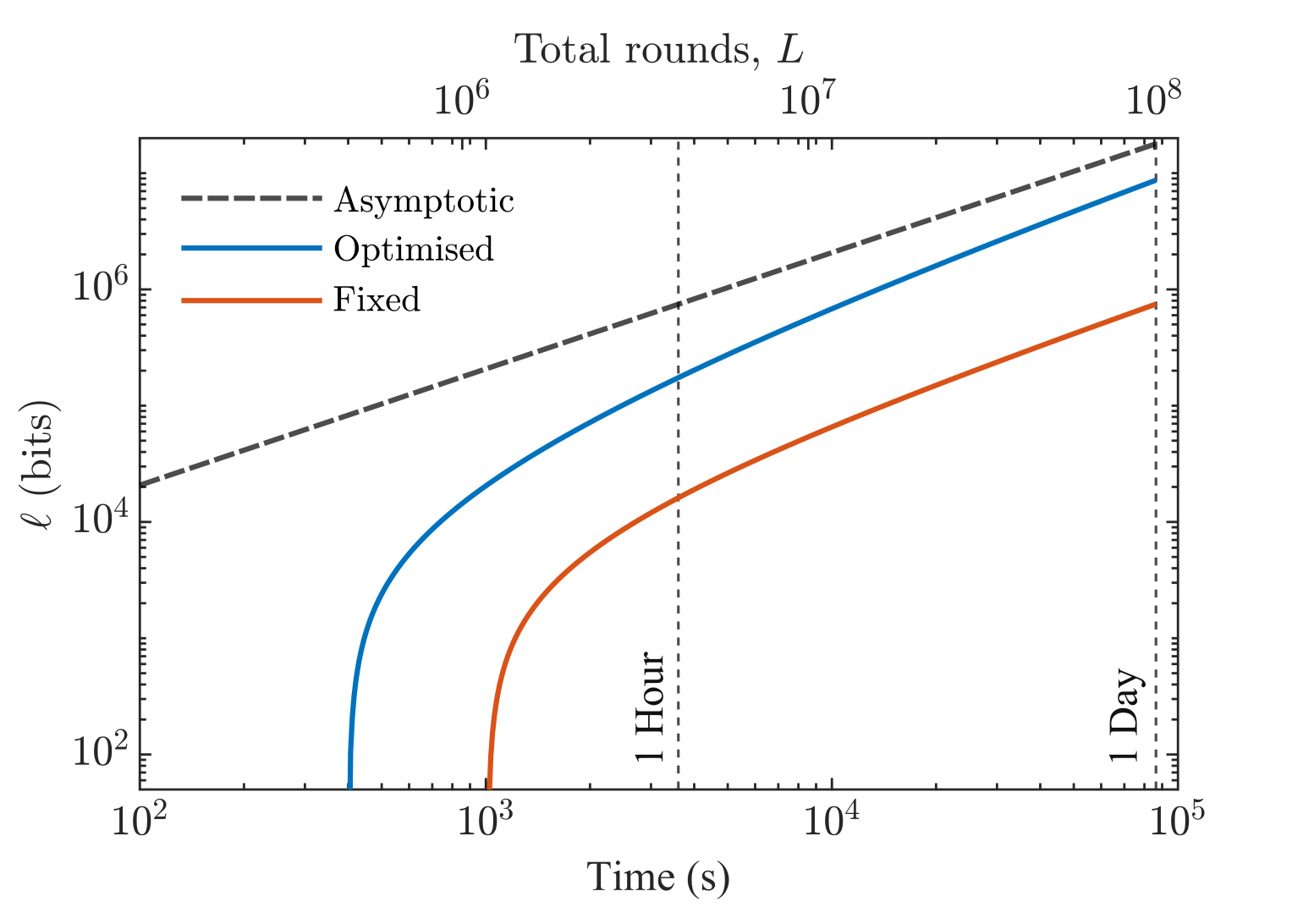}
    \caption{\textbf{Finite key analysis}. 
    Projected extractable secret-key length based on the measured 20-km error statistics and event rate as a function of aggregation time up to 24 hours (bottom axis) and total rounds (top axis).
    The optimised secure key is calculated using the optimal key basis probability $p_{\rm key}^*$, while the unoptimised secure key (fixed) is calculated using a constant $p_{\rm key}=0.5$. The asymptotic secure key is calculated using Eq.~(\ref{eq_akr}).
    A positive key is obtained after $6.5$ minutes, using the optimised basis bias and $4.7\times10^5$ total rounds ($L$).} 
    \label{fig:finite_key}
\end{figure}

The results in Fig.~\ref{fig:finite_key} are obtained by applying the finite-key analysis to the measured data and rescaling the numbers of key and check rounds according to the respective basis probabilities.
Projecting the measured event rates and error statistics to a 24-hour continuous execution of the randomised protocol gives a finite-key lower bound of $7.4\times 10^5$ using the symmetric basis probability, and 8.7~Mbits using the optimised basis bias.

\section{\label{discuss}Discussion}

In bottleneck topologies, the GHZ state can establish correlations between 
all participants in a single network use rather than $n$ links for bipartite protocols~\cite{Epping_2017,pickston2023conference}. 
While the advantage is limited to short distances and a few users, this can be extended to larger networks using quantum repeaters~\cite{memmen2023advantageQSS}. Alternatively, if the dealer owns the source, they can use a post-matching method recently demonstrated in Ref.~\cite{xiao2025experimental} with improved loss scaling. This method establishes multiparty correlations by distributing bipartite states, but requires more photon sources and additional links with the dealer.

In QSS, participants are assumed to be untrusted, which prevents pre-sharing the measurement bases, as in QKD~\cite{erven2009entangled}. Typically, random and symmetric measurement bases are used to protect against participant attacks~\cite{williams2019quantum, kogias2017unconditional}. However, this causes a large basis mismatch and limits the key generation probability to $\eta_{\mathrm{key}}=1/2^n$. Here, following the security analysis of Ref.~\cite{memmen2023advantageQSS}, we can bias the measurement probabilities to optimise key generation rounds and reach the same asymptotic efficiency as trusted QKD protocols. This was proposed in prior work but without composable security~\cite{xiao2004efficient}. Although we implemented the basis bias in post-processing, it can be implemented using passive optics, such as a biased beam splitter~\cite{ribordy2000long} and two additional detectors per user, provided the measurement devices remain trusted~\cite{larsson2014loopholes}.

This experiment made use of bright and high-purity, entangled photon sources to generate GHZ states at up to $5\times 10^3$~(counts/s) achieving a secure shared secret key rate of $>7\times10^2$~bits/sec. 
This is comparable to leading experiments in terms of four-photon GHZ generation rates and quantum state fidelity~\cite{wang2016experimental,zhong201812}, providing a new benchmark for QSS key rates. 
The optimal pump power is relatively high for SPDC sources, resulting in a high multiphoton component.
This can be mitigated by using high repetition rate lasers or time-multiplexed sources to suppress multiphoton emission, while maintaining high generation rates~\cite{broome2011reducing, smirr2011simple, greganti2018tuning}.
Furthermore, \emph{photon-number resolving detectors} (PNRDs) can be used to reject multiphoton errors and allow SPDC sources to operate at even higher rates.
We present an analysis in Appendix~\ref{appen_filter} using pseudo PNRDs, showing that the SKR can increase by up to a factor of three.
However, this method cannot be directly applied using threshold detectors because of a known security loophole~\cite{Lutkenhaus:1999,Lutkenhaus:2000,gottesman2004security,ma2007quantum,Beaudry:2008,Tsurumaru:2008,Gittsovich:2014,bozzio2021multiphoton}.

We now specify the adversarial model and the precise scope of the security claim. We consider a trusted dealer and an arbitrary unauthorised coalition consisting of an external quantum adversary together with up to $n-1$ dishonest participants. The source and quantum 
channels may be fully controlled by this coalition, and attacks may be coherent across all rounds. The measurement devices, private randomness and laboratories of honest users are trusted, and classical communication is authenticated. 
Security means that the protocol is $\varepsilon_{\rm sec}$-secret in the composable trace-distance sense: the trace distance between the actual key-adversary state and an ideal uniform key independent of the adversary's quantum system and the public transcript, weighted by the probability that the protocol does not abort, is at most $\varepsilon_{\rm sec}$.
Dishonest participants may nevertheless force an abort.

The present experiment, which is a proof of principle, differs from a fully cryptographic execution in one important respect.
Owing to the switching time of the motorised analysers, the present experiment records measurement settings in blocks rather than choosing them independently in each round.
The finite-key values should therefore be interpreted as a performance projection for an implementation with genuine random basis selection, under the assumption that the source, channels, and measurement apparatus remain stationary.
Blockwise acquisition by itself does not reproduce the adversarial random-sampling condition required for composable security against general attacks.
A fully cryptographic implementation would require fast active basis selection or an appropriately analysed passive implementation, followed by actual information reconciliation and privacy amplification.

Scaling up this demonstration to larger networks will require near-deterministic sources and entanglement gates due to the unfavourable scaling of SPDC sources with the number of users. Progress has been made in the design of deterministic emitters, with quantum dot sources already demonstrating heralded 3-photon GHZ generation at approximately 500 (counts/s), although current fidelities remain around 0.72~\cite{Pwalther24}. Alternatively, probabilistic sources can be time multiplexed using fast optical switching to approximate deterministic operation~\cite{Silberhorn12}. Meanwhile, boosted fusion schemes are reducing the gate overhead, with recent experiments reporting a 0.71 fusion success probability~\cite{boost_fusion26}. Continuous variable implementations have further extended QSS, pushing transmission distances to 55~km~\cite{liu2023experimental} and supporting networks of up to 24 participants~\cite{zhang2026large}.

In this work, we have considered secret sharing of classical information with quantum security. A natural next step is to extend this framework to quantum state sharing~\cite{lance2004tripartite,lu2016secret}, which would connect QSS with distributed quantum computing, quantum error correction and consensus protocols~\cite{broadbent2009universal,M.Anne13}. These tasks may require different access structures than the symmetric $(n,n)$-threshold scheme investigated in this work. To make use of the general $(k,n)$-threshold connectivity available with graph states~\cite{markham2008graph,bell2014experimental} will require extending composable security proofs to these graph state schemes. This will enable protection against dishonest participants with flexible access structures and substantially broaden the range of quantum network applications.
We hope that this work stimulates further practically minded experimental realisations of multipartite quantum cryptographic protocols, accompanied by transparent and rigorous security analyses.

\begin{acknowledgments}
We acknowledge support from the UK Engineering and Physical Sciences Research Council (Grant Nos.~EP/Z533208/1). J.~M.\ and A.~P.\ acknowledge funding from the Quant-GPlCz Project (supported by BMDV grant 
No.~19OS25001A and the EU’s Horizon Europe Programme grant 
No.~101249338), the QIA Phase 1 project (supported by the European Union's Horizon Europe research and innovation programme under grant agreement 
No.101102140) and Berlin Quantum Alliance.
J.~E. and N.~W.
acknowledge support from the BMFTR (QR.N), Berlin Quantum, the Munich Quantum Valley, the DFG (SPP 2514, CRC 183), the Clusters of Excellence (MATH+ and ML4Q),
and the European Research Council (DebuQC). For Berlin Quantum, this is the result of a joint-node collaboration.

\end{acknowledgments}

\appendix

\section{Experimental materials}\label{supp_exp}

In the main text, we have briefly outlined the GHZ preparation. Here we provide more detail on the source setup and characterisation. We used a pulsed Ti:sapphire laser operating at 80~MHz repetition rate, 1.3~ps pulse duration, and central wavelength at 775~nm to pump the photon-pair sources. Each source undergoes type-II spontaneous paramteric down conversion, generating photon pairs at approximately 1550~nm with opposite-oriented polarisation. When placed inside a Sagnac interferometer, the sources are pumped in counter-propagating directions, which creates entanglement in the polarisation degree of freedom through path erasure. The photons are coupled into a single-mode fibre (SMF-28) and sent to polarisation analysers. We prepare the $\ket{\Phi^{+}}$ Bell state vector given by
\begin{equation}
    |\Phi^{+}\rangle=\frac{1}{\sqrt{2}}\Big(|h,h
    \rangle + |v,v\rangle\Big),
\end{equation}
using \emph{fibre polarisation controllers} (FPC) 
and optimise the contrast between eigenvalues in the $\langle ZZ\rangle$ and $\langle XX\rangle$ bases.
The photonic state is characterised using full quantum state tomography and maximum-likelihood estimation to reconstruct the density matrix. 
At 20~mW pump power, we measure a Bell state fidelity of $0.991\pm 0.003$ and $0.990\pm 0.003$ for each source, respectively. Extrapolating the power dependence, we find the Bell state fidelity at 600~mW is $0.928\pm 0.005$ for both sources.  

Before interfering the independent sources at the fusion gate, each source was first tuned to spectral degeneracy using dependent two-photon interference. The signal and idler wavelengths were controlled by a crystal oven, and the temperature was swept until the HOM visibility was maximised. This provides the wavelength matching condition at 1550~nm, since no tight spectral filtering is used. Instead, the domain-engineered sources suppress side lobes in the joint spectrum, with only silicon long-pass windows used to remove residual pump light. The interference fringes are measured in the joint $X$  basis $\langle XX \rangle$. At 20~mW pump power, the two sources showed dependent two-photon interference visibility of $0.981\pm0.002$ and $0.992\pm 0.002$, respectively, which confirms a high overlap in the spectral component.

Next, we characterise the fusion gate using interference between photons from independent sources. In this measurement, successful events are post-selected on 4-fold coincidences. The resulting two-photon interference visibility provides an operational indicator of the source purity and residual distinguishability in the spectral, temporal and polarisation modes~\cite{branczyk2017hong}. The fringes are measured in the joint $X$  basis $\langle XXXX \rangle$.
At 20~mW, we measure an independent two-photon interference visibility of $0.960\pm 0.003$.  When we increase the pump power to 600~mW, visibility falls commensurately to $0.837 \pm 0.003$.

\section{Tomographic recovery}\label{supp_tomo}

After characterising the fusion gate, we prepare the photons in the correct polarisation state using the FPC's. We perform full quantum state tomography on the four-qubit state, measuring all combinations of the joint Pauli basis. There are 81 settings in total, each setting recorded for 30 seconds of data collection. The single-qubit measurement station uses a dual-detector setup with quantum efficiencies ranging from 0.7-0.8. To partially mitigate polarisation-dependent loss from detector imbalance, the detectors monitoring the \emph{transmission}  (T) and \emph{reflection} 
(R) PBS ports are swapped halfway through the measurement. This is achieved by flipping the wave plates to the conjugate basis and swapping the detector labels from $\mathrm{T} \rightarrow \mathrm{R}$. Each setting is recorded for 15 seconds, providing 30 seconds of total integration time. This provides a partial balancing of detector loss. A full compensation would require 16 detector flips per basis using every party combination, but we avoid this due to time constraints. 

\begin{figure}
    \centering
    \includegraphics[width=1\linewidth]{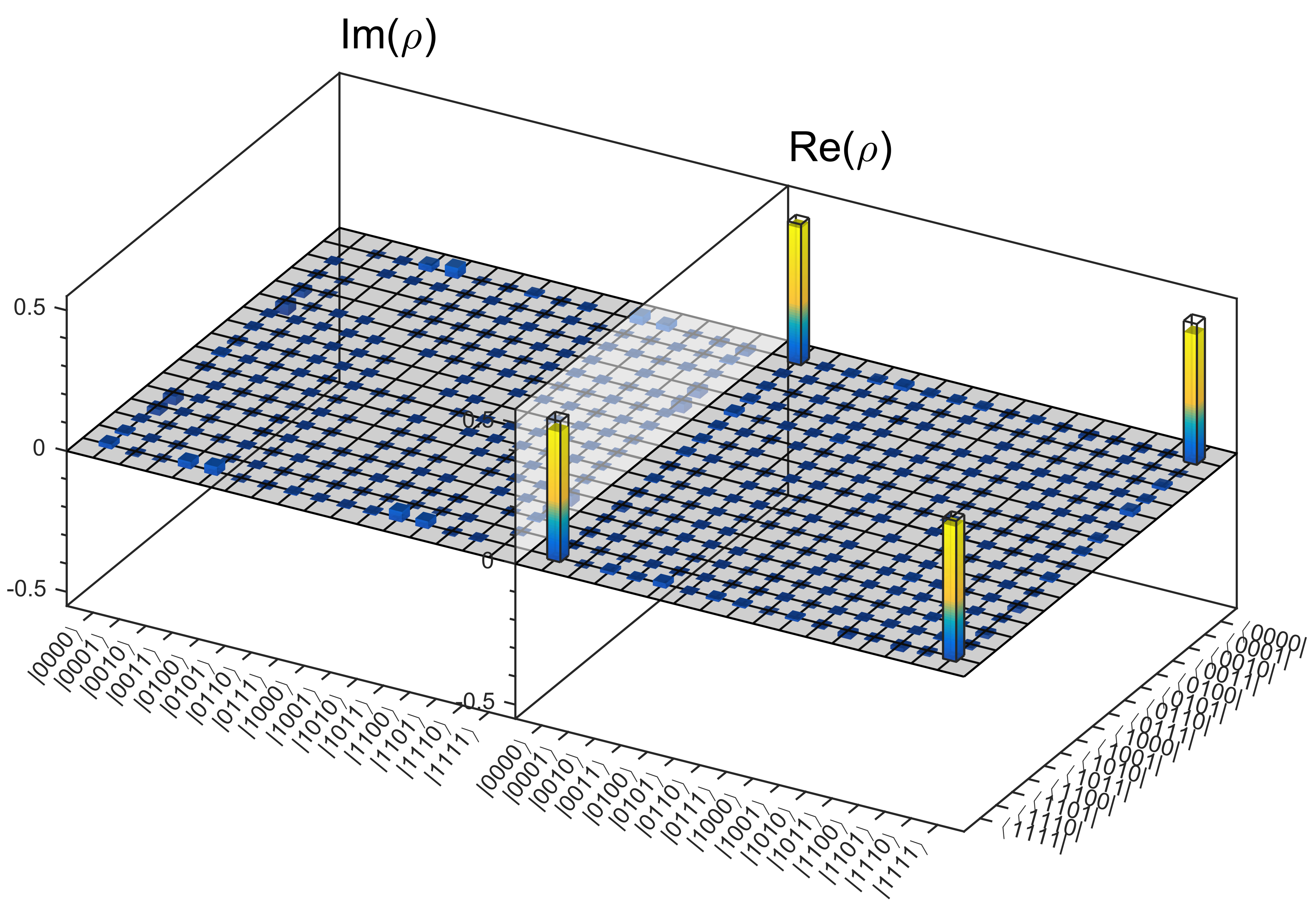}
    \caption{\textbf{Density matrix.} The tomographically reconstructed density matrix of the four-qubit GHZ state prepared using 100~mW pump power and a generation rate of 112 four-folds/s. Each basis is measured for 30 seconds, and we collect $3.3\times10^3$ rounds per basis and $2.7 \times10^5$ rounds in total. The reconstructed density matrix has a state fidelity of $0.948\pm 0.018$ with the ideal GHZ state, shown with the empty bars and a purity of $0.919 \pm 0.013$.}
    \label{fig:tomo}
\end{figure}

\begin{figure}
    \centering
    \includegraphics[width=1\linewidth]{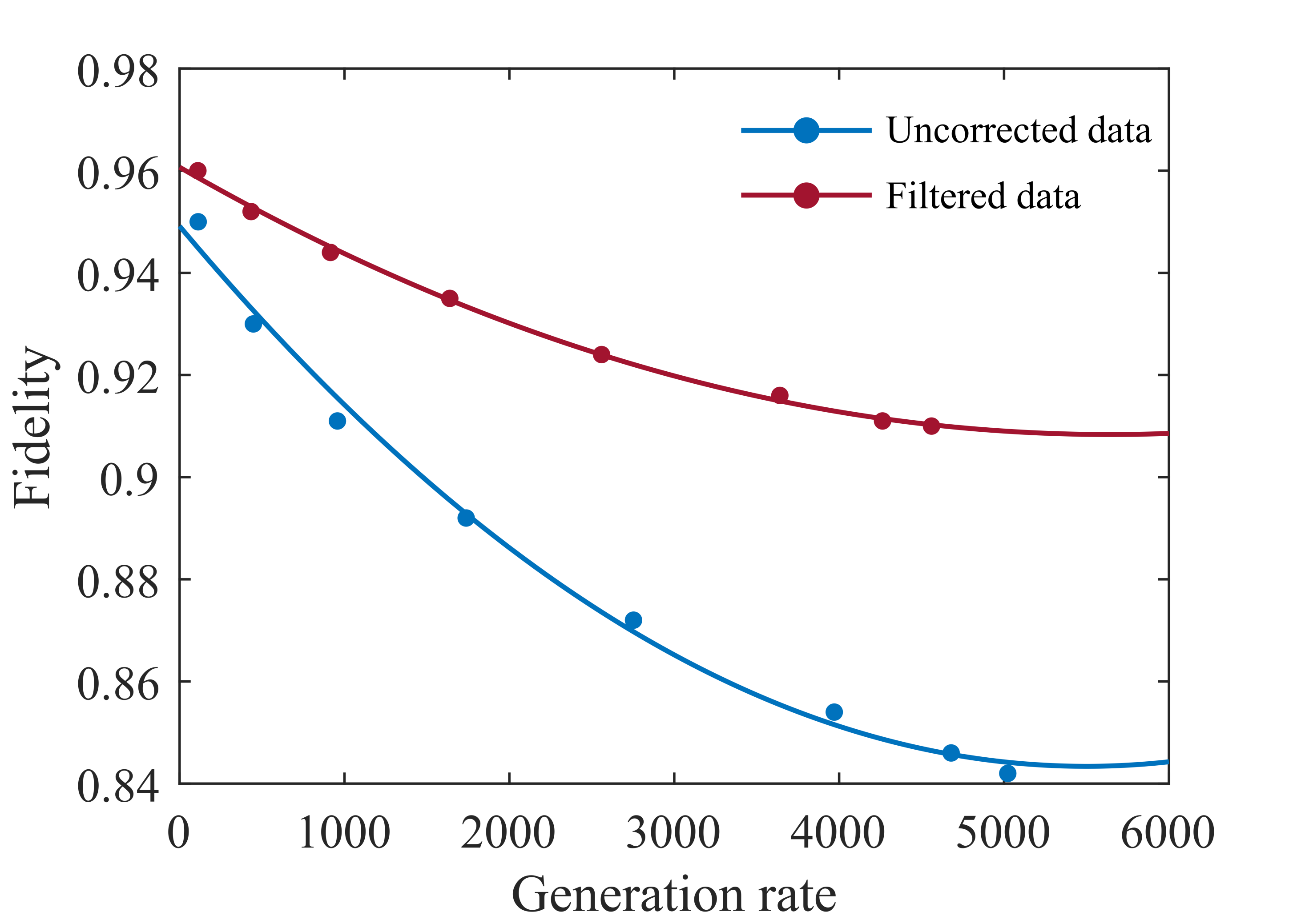}
    \caption{\textbf{GHZ characterisation}. The quantum state fidelity of the experimental GHZ state is plotted against the generation rate (four-folds/s). In blue we keep all double-click events, while in red, we discard these events, see Appendix~\ref{appen_filter}. Each basis is measured for 30 seconds at all power levels. Error bars have been calculated and excluded due to the size.}
    \label{fig:fid_pump}
\end{figure}

The density matrix is reconstructed using maximum-likelihood estimation and presented in Fig.~\ref{fig:tomo}. Statistical uncertainties are obtained from the standard deviation of 100 Monte Carlo simulations assuming Poisson photon statistics, but are relatively small and error bars are neglected. We calculate the quantum state fidelity using the Uhlmann fidelity,
\begin{equation}
    F(\rho, \tau) = \bigg(\mathrm{tr}\sqrt{\sqrt{\rho} \tau \sqrt{\rho}}\bigg)^2,
\end{equation}
where $\rho$ is the reconstructed density matrix and $\tau=|\rm GHZ\rangle \langle \rm {GHZ}|$ is the ideal GHZ state. We repeat the tomography at a range of pump powers and plot the corresponding fidelity and generation rate in Fig.~\ref{fig:fid_pump}. At maximum power of 680~mW, we obtain a fidelity of $0.842 \pm 0.002$ with a generation of 5025 (counts/s). Decreasing the pump power to 100~mW, we obtain a fidelity of $0.948 \pm 0.018$ and a generation rate of 111 (counts/s).

\section{Source optimisation}\label{app_optimal}

We have evaluated the effect of pump power on the state fidelity and generation rate; now we analyse the effect on the \emph{secure key rate} (SKR).
The SKR is a function of both the generation rate and the error rate, which are plotted in Fig.~\ref{fig:optimisation} with the source pump power. It can be seen that the generation rate increases quadratically, while the error rates increase linearly with pump power. The \emph{asymptotic key rate} (AKR) is calculated using Eq.\ (1) in the main body and also displays a linear relationship with pump power. At around 800~mW the AKR falls to zero where no positive key can be obtained. When we multiply the generation rate $r$ by the AKR we find the SKR plotted in Fig.~\ref{fig:optimisation}(d). This results in a non-linear behaviour of the SKR, with a peak at 550~mW in the fitted curve. This identifies the optimum pump power for our source for QSS with the maximum key rate.  The optimal experimental power level used was 600~mW due to limitations in setting different wave plate angles using motorised rotation stages.

\section{Multi-photon filtering}\label{appen_filter}

Multiphoton emission is a fundamental limitation of SPDC sources and becomes proportionally larger at higher pump powers.  
A type-II SPDC source emits the two-mode squeezed vacuum state~\cite{ma2007quantum} vector given by
\begin{equation}
    \ket{\Psi}=(\mathrm{cosh}\chi)^{-2} \sum_{n=0}^\infty \sqrt{n+1}\mathrm{tanh}^n \chi\ket{\Phi_n},
\end{equation}
where \begin{equation}
\ket{\Phi_n}=\frac{1}{\sqrt{n+1}}\sum^n_{m=0}(-1)^m\ket{n-m, m}_a \ket{m, n-m}_b 
\end{equation}
is the $n$-photon pair state vector and $\chi$ is the squeezing term which contains the crystal non-linearity, the pump power and focusing conditions. 

\begin{figure*}
    \centering
    \includegraphics[width=0.95\linewidth]{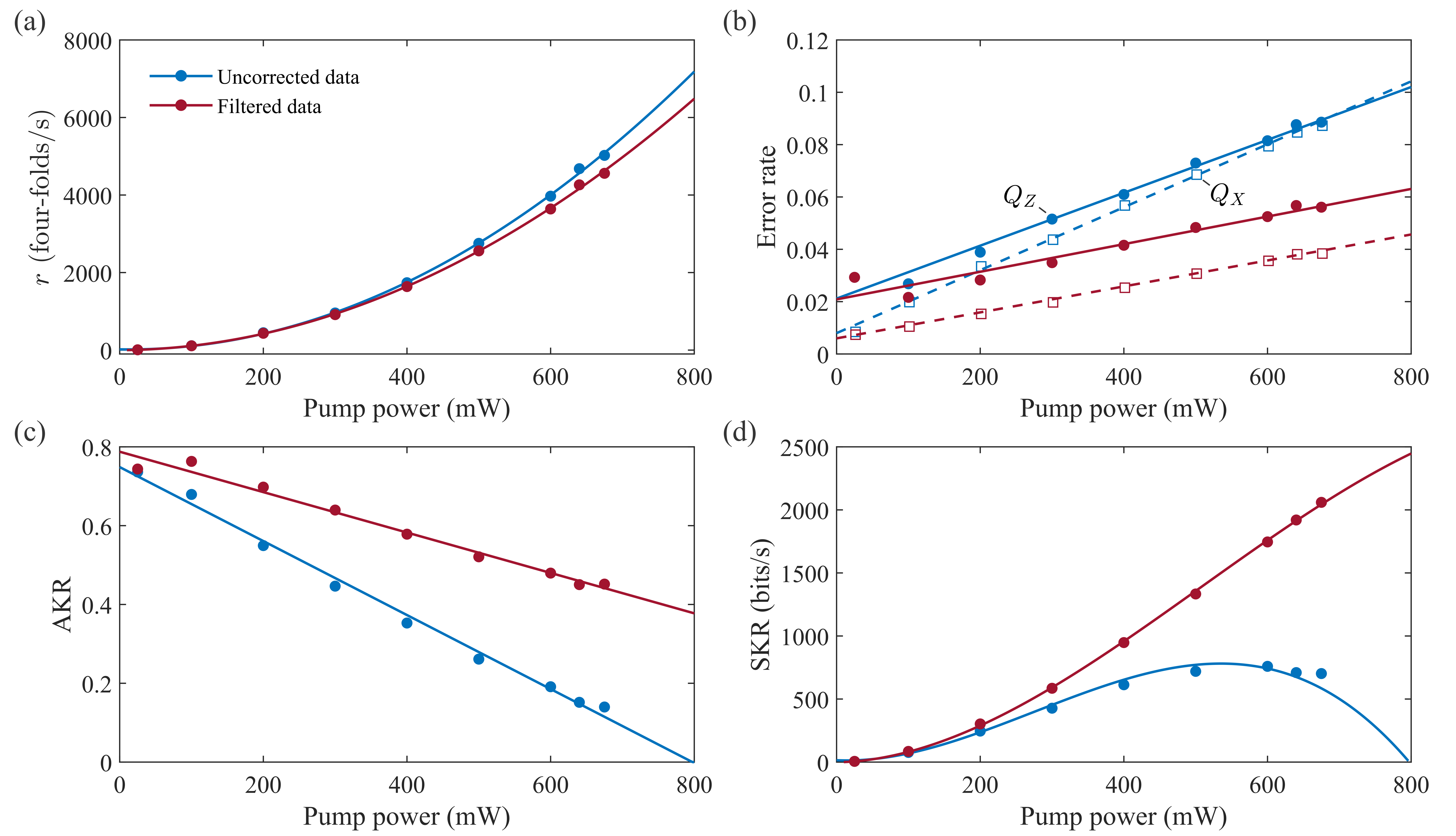}
    \caption{\textbf{Power optimisation}. (a) GHZ generation rate $r$ over a range of source pump power. In blue are the uncorrected data points, where all double-click rounds are kept. In red are the filtered data points which discard double-click events. (b) The quantum bit error rate $Q_z$ and the phase error $Q_x$ are plotted against the source power. (c) The \emph{asymptotic key rate} (AKR) is plotted against the source power. (d) The \emph{secure key rate} (SKR) is plotted against the source power. }
    \label{fig:optimisation}
\end{figure*}

Here, we investigate the effects of multiphoton filtering on the QSS \emph{secret key rate} (SKR).
Our measurement stations are equipped with two detectors that provide pseudo-PNRD capability. When both detectors fire at the same time, this is called a double-click round. Using the fast timing logic, we can filter these events live during data collection. The filtered data has been plotted in Fig.s \ref{fig:fid_pump} and \ref{fig:optimisation}  using red data points. We observe a small penalty in the total generation rate due to the discarded rounds, but a large improvement in the error rate and AKR. For example, at 600~mW the unfiltered $Q_x=0.080$, while the filtered $Q_x=0.035$; a 56\% decrease, and results in an approximate 250\% increase in the AKR. Extrapolating the filtered SKR curve, we find a maximum value of 2700 (bits/s) at 900~mW. The non-linear improvement in SKR is partly due to the Shannon binary entropy $h_2(x):=-x\mathrm{log}_2(x)-(1-x)\mathrm{log}_2(1-x)$ with a non-linear dependence on the error rate $x$.

While the multiphoton filtering demonstrated here provides a 3-fold improvement in the SKR, it cannot be implemented securely in the present experimental setup because of a known side channel attack. In some of the earliest QKD analysis it was pointed out that when using threshold detectors, Eve could control the incident mean photon number and thereby bias which rounds are discarded, potentially revealing or controlling which measurement basis used in the retained rounds \cite{Lutkenhaus:1999,Lutkenhaus:2000}. This issue is particularly acute when a decoy-state protocol (which effectively upper bounds the number of multi photon events) is unavailable, either for practical reasons or because the source is untrusted as is the case here. Fortunately, the security of multiphoton events in this scenario has been extensively studied in Refs.~\cite{Lutkenhaus:1999,Lutkenhaus:2000,gottesman2004security,ma2007quantum,Beaudry:2008,Tsurumaru:2008,Gittsovich:2014,bozzio2021multiphoton}. The only secure approach is to retain all double-click events and assign each a random outcome. This solution was eventually made completely rigorous by showing that a standard threshold detector provably admits a so-called squashing model \cite{Beaudry:2008,Tsurumaru:2008,Gittsovich:2014}. By contrast, if photon number resolving detectors were placed after the polarisation measurement, the photon number information would be independent of the measurement basis, allowing the filtering procedure to be applied securely.

\section{Finite key analysis}\label{supp_finite}

\begin{table}[t]
\centering
\small
\begin{tabular}{lcc}
\hline
Description \hspace{1em} & Parameter \hspace{1em} & Value \hspace{1em}\\
\hline
Failure probability  & $\varepsilon$ & $10^{-10}$ \\
Robustness  & $\varepsilon_{\mathrm{rob}}$ & $\varepsilon/8$ \\
Error correction & $\varepsilon_{\mathrm{EC}}$ & $\varepsilon/2$ \\
Privacy amplification  & $\varepsilon_{\mathrm{PA}}$ & $\varepsilon/4$ \\
Parameter estimation  & $\varepsilon_{\mathrm{PE}}$ & $\varepsilon/8$ \\
\hline
\end{tabular}
\caption{\textbf{Finite key constants}. The budget for a $\varepsilon$-secure protocol is $\varepsilon= \varepsilon_{rob} + \varepsilon_{\mathrm{EC}}+\varepsilon_{\mathrm{PA}}+\varepsilon_{\mathrm{PE}}$. For convenience, we have used $\varepsilon_{\mathrm{rob}}=\varepsilon_{\mathrm{PE}}$.}
\label{table_params}
\end{table}

The finite-size composable secure key uses entropic uncertainty relations to bound the information obtained by the dishonest parties, given by;
\begin{equation}\label{entrop_rel}
    \ell = \mathrm{min}_j H_{\mathrm{min}}^{\varepsilon_{\mathrm{PE}}/p_{\mathrm{p}}} (\boldsymbol{\mathrm{X}}_{A}^m|E,~U_j) - \ell_{\mathrm{EC}} - \mathrm{log}_2 \bigg(\frac{1}{4\varepsilon^{2}_{\mathrm{PA}}} \bigg).
\end{equation}
We will summarise the core components, while the complete derivation can be found in Ref.~\cite{memmen2023advantageQSS}. The first term in Eq.~(\ref{entrop_rel}) is the eavesdroppers ($E$) smooth conditional min-entropy of the random variable $\boldsymbol{X}^m$, given the untrusted set of Bob's $U_j$, generated by $m$ measurements of the key-generating observable. The exponent contains $\varepsilon_{\mathrm{PE}}$ the security parameter for parameter estimation, see Table~\ref{table_params}, as well as $p_{pa}$, the probability that the protocol passes all parameter estimation and error correction tests and does not abort. The second term corrects for the information leaked during error correction, and the last term accounts for the fraction of the sifted key used for privacy amplification. 

The operational form of Eq.~(\ref{entrop_rel}) is used in Eq.~(\ref{fin_key}) to evaluate the secure key length for the 20~km fibre experiment. The $\epsilon$ security parameters used are given in Table~\ref{table_params}. We consider both an unbiased basis choice, with $p_{\mathrm{key}}=0.5$, and an optimised basis probability $p_{\rm key}^*$. The number of key generation rounds is given by $m=\eta_{\mathrm{key}}L$, where $\eta_{\mathrm{key}}=p_{\mathrm{key}}^4$ is the probability that all 4 users select the key basis. Similarly, the probability of obtaining a valid check round is $\eta_c=(1-p_{\mathrm{key}})(1-p_{\mathrm{key}}^{n})$.

The experiment used motorised wave plate stages to perform independent basis selection. However, the rotation times are too long to make a change on every round and are impractical during a long finite key acquisition. We therefore measured the relevant $X$ and $Z$ basis combinations for equal durations, providing uniformly sampled estimates of the protocol error rates. The dataset was collected in 200 blocks, each containing approximately 110,000 rounds. Each block included the 12 measurement settings contributing to either the key rounds $m$ or the check rounds $k_i$, while the remaining 4 settings were omitted. In total, more than $22$ million rounds were recorded.

\begin{figure}
    \centering
    \includegraphics[width=0.95\linewidth]{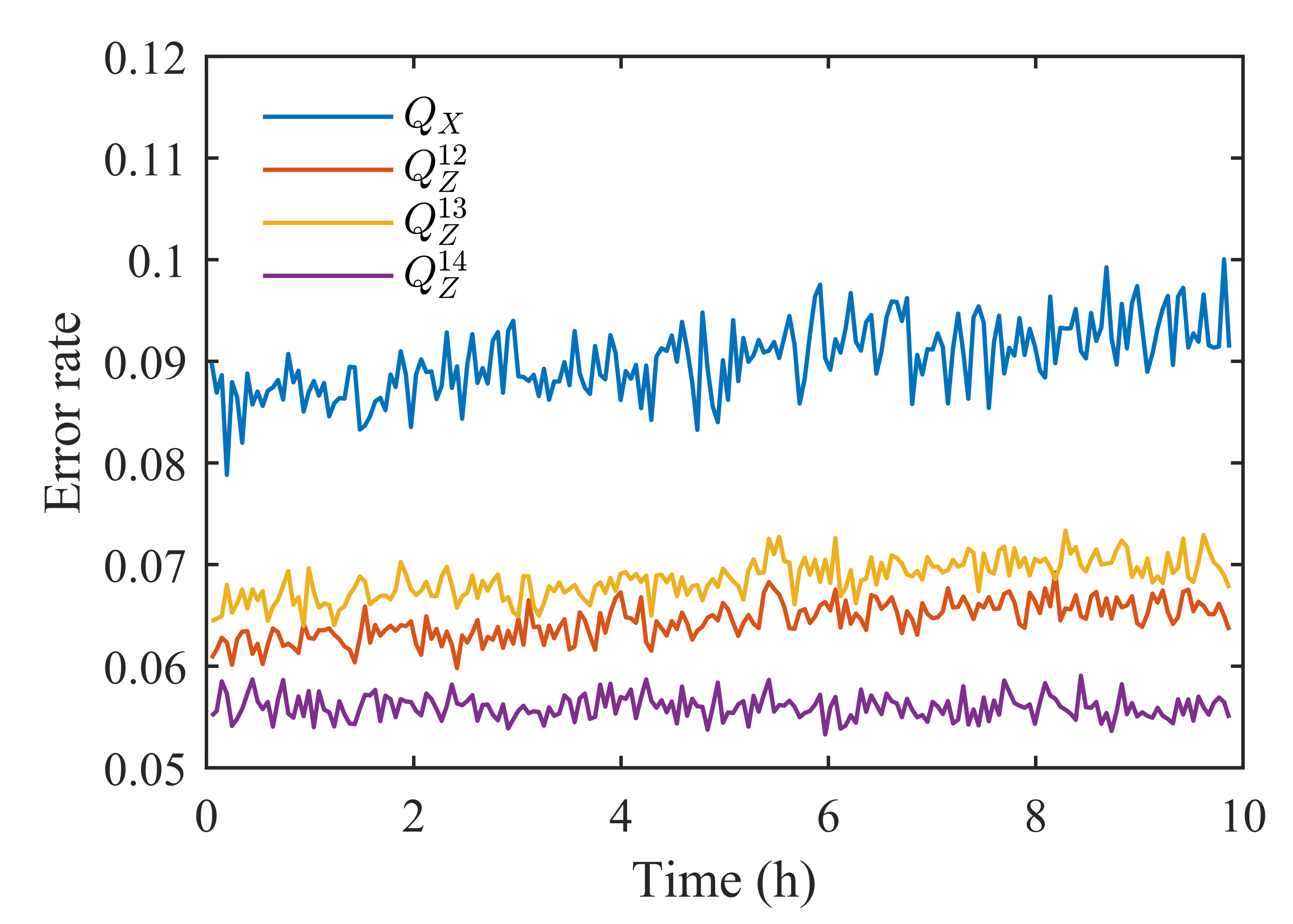}
    \caption{\textbf{Error tracking}. The error rates through 20~km fibre over a 10~hour collection time. Both the joint $X$  error $Q_X$ and the pairwise $Z$  errors $Q_Z^{i,j}$ are plotted for the best dealer, player 1. Each time series is obtained from 200 measurements, equally spaced over the entire duration. Each measurement block contains equally weighted $Z$  and $X$  bases to allow all combinations of calculations using 8 seconds of measurement time.}
    \label{fig:aggre_err}
\end{figure}

\begin{figure}
    \centering
    \includegraphics[width=1\linewidth]{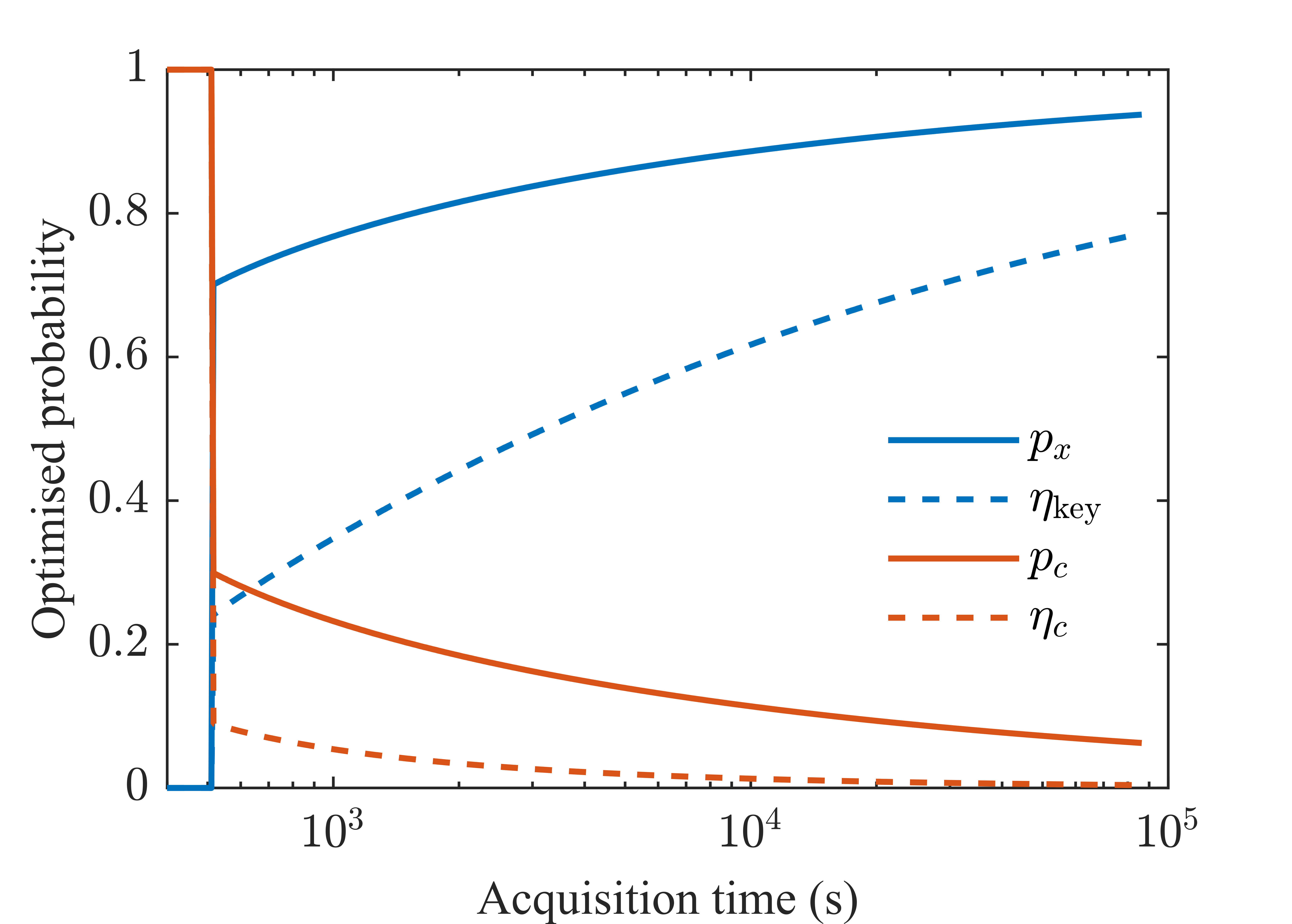}
    \caption{\textbf{Measurement basis optimisation}. The optimum measurement basis probability for the experimental QSS demonstration through 20~km of fibre. The initial bias is set to $p_x=0$ and the check basis probability $p_c=1-p_x$. }
    \label{fig:basis_opt}
\end{figure}

The secure key length was then evaluated in post-processing using the measured cumulative values of $Q_X$ and $Q_{A,B_i}$ plotted in figure~\ref{fig:aggre_err}. For each aggregation length of raw key $L$, the basis bias $p_{\mathrm{key}}$ was optimised to maximise Eq.~(\ref{fin_key}), the values have been plotted in Fig.~\ref{fig:basis_opt}. The optimum key basis bias approaches $p_{\rm key}^*=0.947$ for the complete dataset.

\section{Network emulation}\label{fibre_net}

Here we discuss how the fibre network is set up to demonstrate the QSS protocol. We used four 5~km single-mode fibre spools (SMF-28) with measured losses between 0.19-0.21 dB/km. The GHZ source is placed in the centre of the network, with each participant connected to the source through the fibre links in a star topology. The fibres are kept in an insulating box to shield them from thermal fluctuations in the lab, which induce random phase shifts and increase the X-basis errors. The fibres all have \emph{fibre polarisation controllers} (FPC) 
installed to correct for random unitary rotations on the transmitted photons. We correct for this by sending a known reference state through the fibres and measuring the output on the polarisation analysers; the fibre unitary is reversed using the FPC's.  

\bibliography{bib.bib}

\end{document}